# The Science Case for the 4π Perspective: A Polar/Global View for Studying the Evolution & Propagation of the Solar Wind and Solar Transients

**Submitted by:** *Angelos Vourlidas, S. Gibson, D. Hassler, T. Hoeksema, M. Linton, N. Lugaz, J. Newmark*

**Endorsed by**: *B. Alterman, N. Arge, C. Braga, G. Chintzoglou, C. Downs, K-Y. Ko, E. Mason, S. di Matteo, M-P. Miralles, T. Nieves-Chinchilla, R. Nikoukar, L. Rachmeler, Y. Rivera, G. deToma*

## Solar Wind and Transients: Current State

STEREO revolutionized heliospheric physics by providing consistent, spatially resolved imaging of the inner heliosphere with coverage from the Sun to 1 AU. Thanks to their off- Sun-Earth Line (SEL) vantage points, the STEREO coronal and heliospheric imagers can *uninterruptedly* observe Corotating Interaction Regions (CIRs) and Coronal Mass Ejections (CMEs) en route to Earth and image directly the solar wind flow (**Fig. 1**). STEREO's novel dual-viewpoint coronal observations revolutionized solar physics. They enabled the 3D reconstructions of coronal loops, CMEs, shocks and CIRs and led to the first global (360°) imaging of the Sun in the corona (EUV and coronagraphs) and first global view of the inner heliosphere in the ecliptic with heliospheric imagers. The observations uncovered the magnetic flux rope nature of CMEs, mapped the speed and density profiles across shocks, and revealed an unexpectedly wide longitudinal spread of SEPs in the heliosphere. The global coronal imaging provided some tantalizing evidence for long-distance connections in solar eruptivity. Additional viewpoints from SOHO and SDO, extended the studies of transient evolution and interactions, naturally feeding an increasingly sophisticated space-weather research field and demonstrating the power of a systems-approach for advancing science and operations.

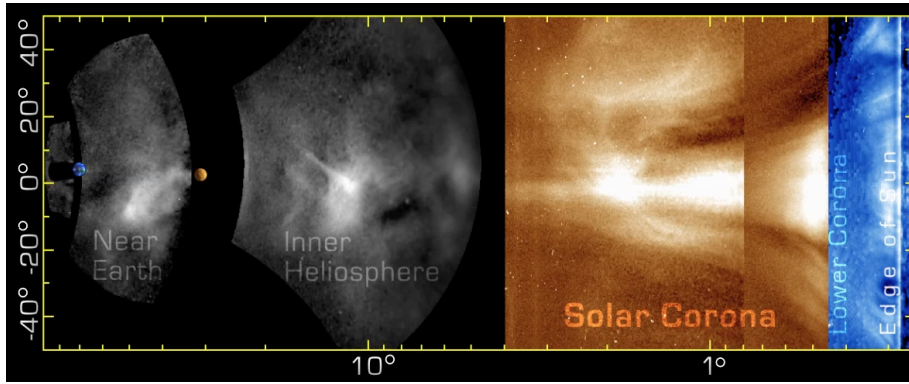

*Fig. 1*: *STEREO is the first mission to observe uninterruptedly the inner heliosphere from Sun to Earth as the composite of the STEREO-A imagers demonstrates (From DeForest et al., 2013)*

## Solar Wind and Transients: Desired State

While the STEREO mission demonstrates that we can indeed image the solar wind and its major components, it has left us with many unanswered questions and a foreshortened 360° view of the poles. With regards to CMEs, heliospheric imaging analyses provide some indications of CME rotational evolution en route to Earth (e.g., Isavnin+2014) and multiple cases of CME-CME interactions (Lugaz+2017, and references therein), but it is hard to establish the reliability and general applicability of these results due to projection effects, long lines of sight through overlapping structures, and potentially complex internal morphology. Evolutionary effects such as ambient plasma pileup can also modify the brightness distribution around the transient, and skew triangulation or 3D reconstruction results. This is particularly relevant to longitudinal deflections (Wang+ 2004; Isavnin+ 2014) and widths (of major importance to the interpretation of SEP characteristics, e.g. Lario+ 2017), as their study relies almost exclusively on imaging from within the ecliptic. The wide coverage of the STEREO/EUVI + SDO/AIA imagers allowed detailed studies of CME source regions and their evolution for their full lifetimes for the first time but...only for a few years (2010-14). The exploration of long-range interactions, identification of 'stealth' CMEs and a complete understanding of how magnetic energy flows from the interior to the corona, how it is stored and then explosively released cannot be accomplished by single viewpoint observations. To continue the pathfinding STEREO studies, we need to **re-establish the 360º coronal (and photospheric magnetic field) coverage from the ecliptic** (**Fig. 2**).

While re-establishing the 360º view is an obvious next step, we need to think well beyond that by developing high latitude (>60º) observations to make fundamental leaps in Heliophysics and Space Weather understanding[1]. The STEREO imagers most efficiently resolved the 3D structure of transients only at separations above 60º. **A polar viewpoint,** unavailable even to the Solar Orbiter mission which will reach only 34° above the ecliptic**, comprehensively visualizes both the radial and longitudinal evolution of the solar wind and transient structures propagating through it** (**Fig. 2**)[2]. It is particularly good for capturing the formation and evolution of CIRs and shocks in a continuous fashion, for investigating the magnetic connectivity across large swaths of space, and for following far- and Earth-side surface activity *simultaneously*[3], thus enabling linking small and large-scale activity (e.g flux emergence and global coronal reconfiguration) to solar eruptivity. It may be the only way to determine whether eruptive activity is influenced by long-range interactions, i.e. 'sympathetic' CMEs.[1] The imaging of the corona and inner heliosphere, along with their embedded transients, *from the solar rotation viewpoint* will give us, for the first time, the actual global view of the system we are so accustomed to in our models (**Fig.2, right**) allowing us to see *all the solar-wind transients, all the time*. For maximum science return, these should be coupled with complementary SEL/non-SEL in-situ measurements during CME-CME or CME-CIR interactions.

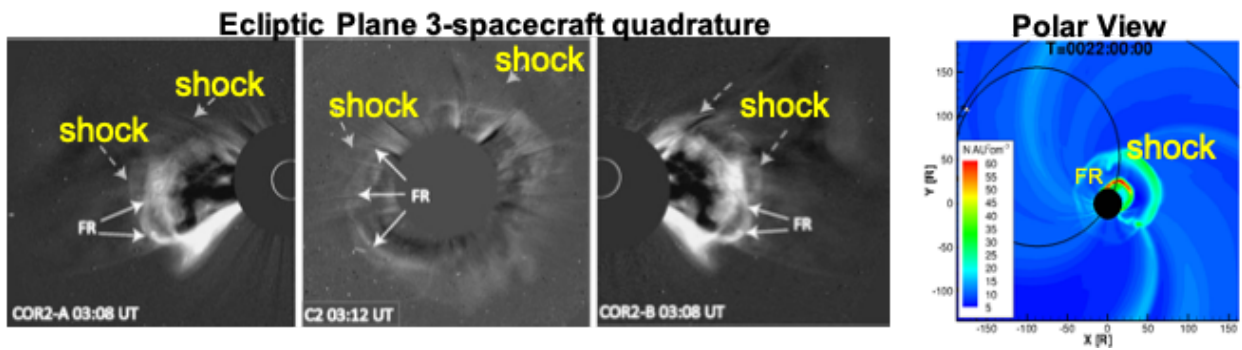

*Fig. 2*: Left panels: Earth-directed CME imaged by 3 coronagraphs in quadrature. The shock sheath and the magnetic flux rope (FR) driver can be discerned but the true extent of the shock and accurate CME directions can only be obtained by a polar perspective (right panel). Adapted from Vourlidas et al, 2013; MHD simulation courtesy of N. Lugaz.

Naturally, we will reap the maximum benefits from a systems approach, incorporating both polar and ecliptic viewpoints to achieve a true 4π coverage of the solar atmosphere (Gibson+2018; Vourlidas+2018). **Table 1** outlines a notional roadmap for achieving this challenging, yet unparalleled accomplishment in all of Astrophysics.

---

[1] See white paper 'The Science Case for a 4π Perspective: A Polar/Global View on Space Weather Magnetic Origins' by Gibson et al
2 See white paper 'The Science Case for a Polar Perspective: Discovery Space' by Hassler et al
3 See white paper 'The Science Case for a 4𝛑 Perspective: A Polar/Global View for Understanding the Solar Cycle', Hoeksema et al.

*To make progress on the open questions on CME/CIR propagation, their interactions and the role and nature of the ambient solar wind, we need spatially resolved coverage of the inner heliosphere -- both in-situ and (critically) imaging -- at temporal scales matching the evolutionary timescales of these phenomena (tens of minutes to hours), and from multiple vantage points (Gibson et al. 2018). The polar vantage is uniquely beneficial because of the wide coverage and unique perspective it provides (Table 1).*

**Table 1. Notional Timeline for Reaching Closure on CME/CIR Evolution and Propagation Issues**

| **Open science question** | How do transients evolve and interact with the ambient solar wind as they move through the heliosphere? | | | |
|---|---|---|---|---|
| **Objectives** | How CMEs affect the large-scale corona? <br> What is the 3D morphology of CMEs/CIRs/plumes? <br> How do solar wind structures propagate & interact? <br> What are the SpW conditions in the inner heliosphere? | | | |
| **Measurements needed** | (1) White light/multiwavelength imagers/coronagraphs; (2) Heliospheric imagers; (3) in-situ solar-wind measurements | | | |
| | | | | |
| **Benefits from non-SEL vantage** (assumes existence of complementary SEL observations) | | **Polar** | **Quadrature (Ecliptic)** | **Far-side** |
| More lines of sight to reconstruct 3D solar-wind structures | | Yes (1),(2) | Yes (1),(2) | no |
| Complementary in-situ probing and remote imaging of solar wind/transient/SEP evolution | | Yes (1),(2) | Yes (1),(2) | no |
| Longitudinal structure of Alfven surface and of CMEs, CIRs, and shocks revealed | | Yes (1),(2),(3) | partially | no |
| Long-range interactions (i.e. 'sympathetic' CMEs) | | Yes (1) | partially | partially |
| Single-view determination of CME ecliptic direction | | Yes (1),(2) | no | no |
| | | | | |
| **Science Investigation Timeline** [assumes existence of complementary SEL observations] | <2030 | Preliminary exploration of Solar Orbiter high latitude data (2027+) CME/CIR/SEP IP evolution studies (PSP-SO-Bepi) Re-establish STEREO-like ecliptic coverage for SC 25-26 (e.g. $L_5/L_4$), polar pathfinder | | |
| | >2030 | First polar (>60°) investigations, 2-3 view ecliptic system | | |
| | 2050 | 4π coverage of the corona/inner heliosphere | | |